\documentclass[prb,twocolumn,showpacs]{revtex4}
\pdfoutput=1
\usepackage{amsfonts} % collection is a set of miscellaneous TeX fonts
\usepackage{amsmath}
\usepackage{amssymb}
\usepackage{bm}% bold math
\usepackage[usenames, dvipsnames]{color} % use colored tex
\usepackage{dcolumn}
\usepackage{epic} % enhanced picture environment
\usepackage{epsfig}
\usepackage{eucal} % The package changes the font that is selected by \mathcal
\usepackage{graphicx}
\usepackage[breaklinks,colorlinks = true,linkcolor = blue,urlcolor=blue,citecolor=blue]{hyperref}
\usepackage{soul}
\usepackage{subfigure}
\usepackage{ulem}
\usepackage{wrapfig} % to include figure with text wrapping around it
\usepackage{xy} % general drawing package useful for drawing complex commutative diagrams

\newcommand{\beq}{\begin{equation}}
\newcommand{\eeq}{\end{equation}}
\newcommand{\beqa}{\begin{eqnarray}}
\newcommand{\eeqa}{\end{eqnarray}}

\begin{document}

\title{Quantum Phase Transition in Heisenberg-Kitaev Model}
\author{Robert Schaffer$^1$}
\author{Subhro Bhattacharjee$^{1,2}$}
\author{Yong Baek Kim$^{1,3}$}
\affiliation{$^1$ Department of Physics and Center for Quantum Materials, University of Toronto, Toronto, Ontario M5S 1A7, Canada.\\
$^2$ Department of Physics and Astronomy, McMaster University, Hamilton, Ontario  L8S 4M1, Canada.\\
$^3$ School of Physics, Korea Institute for Advanced Study, Seoul 130-722, Korea.}

\date{\today}
%%%%%%%%%%%%%%%%%%%%%%%%%%%%%%%%%%%%%%%%%%%%%%%%%%%%%%%%%%%%%%%%%%
\begin{abstract}
We explore the nature of the quantum phase transition between a magnetically ordered state with collinear spin pattern and a gapless $Z_2$ spin liquid in the Heisenberg-Kitaev model. We construct a slave particle mean field theory for the Heisenberg-Kitaev model in terms of complex fermionic spinons. It is shown that this theory, formulated in the appropriate basis, is capable of describing the Kitaev spin liquid as well as the transition between the gapless $Z_2$ spin liquid and the so-called stripy antiferromagnet. Within our mean field theory, we find a discontinuous transition from the $Z_2$ spin liquid to the stripy antiferromagnet. We argue that subtle spinon confinement effects, associated with the instability of gapped $U(1)$ spin liquid in two spatial dimensions, play an important role at this transition. The possibility of an exotic continuous transition is briefly addressed.
\end{abstract}

\maketitle
%%%%%%%%%%%%%%%%%%%%%%%%%%%%%%%%%%%%%%%%%%
\section{Introduction}

Theoretical understanding of various quantum spin liquid phases and their possible realization in candidate materials consist one of the most important and interesting questions in modern condensed matter physics. \cite{2002_wen,2008_lee,2010_balents} Much impetus in this field was derived from the discovery, by A. Kitaev, of an exactly solvable spin-$1/2$ Hamiltonian on the Honeycomb lattice with a spin liquid ground state.\cite{2006_kitaev} This Hamiltonian, known as the Kitaev model (see below), has been studied intensively and it has been shown to support a gapped and a gapless $Z_2$ spin liquid with fractionalized excitations. While the gapped phase has abelian anyon excitations, the gapless phase, in the presense of appropriate perturbations, supports non-abelian anyons.\cite{2006_kitaev,2007_feng,2008_chen,2007_baskaran}

Interestingly, it has been shown recently by Chaloupka {\it et al.}\cite{2010_chaloupka} that such a Kitaev model can indeed arise in  layered Honeycomb lattice materials in the presence of strong spin-orbit coupling. In particular, they showed that in certain iridate magnetic insulators, the low energy Hamiltonian for the pseudospin $J=1/2$ iridium moments is given by a linear combination of the antiferromagnetic Heisenberg model ($H_{\rm H}$) and the Kitaev model ($H_{\rm K}$):
\begin{align}
H = (1-\alpha )H_{\rm H} - 2\alpha H_{\rm K}
\label{eq_hkhamiltonian}
\end{align}
where $\alpha$, expressed in terms of the microscopic parameters, determines the relative strength of the Heisenberg and the Kitaev interactions (the detailed forms of $H_{\rm H}$ and $H_{\rm K}$ are given below).

\begin{figure}
\centering
\subfigure[]{
\includegraphics[scale=.7]{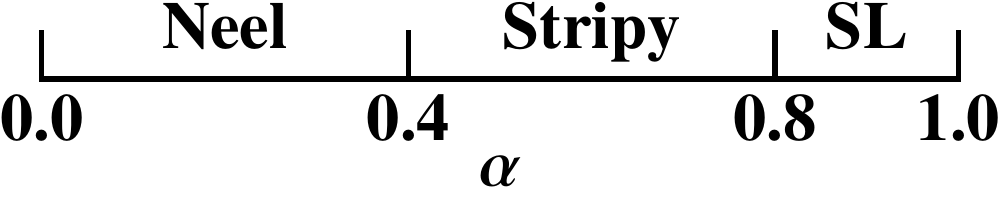}
\label{fig_phasediagram}
}
\subfigure[]{
\includegraphics[scale=.175]{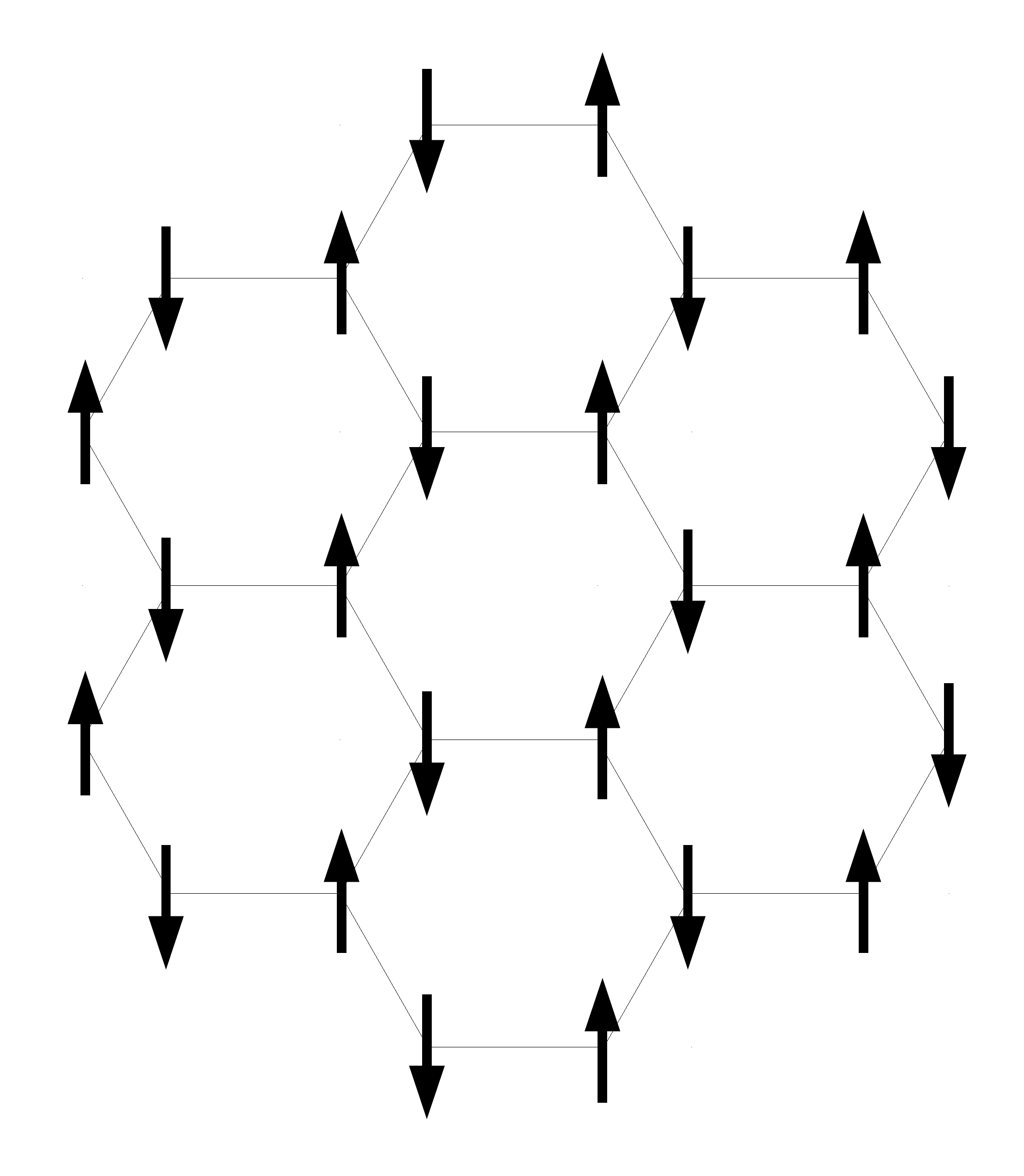}
\label{fig_neel}
}
\subfigure[]{
\includegraphics[scale=0.175]{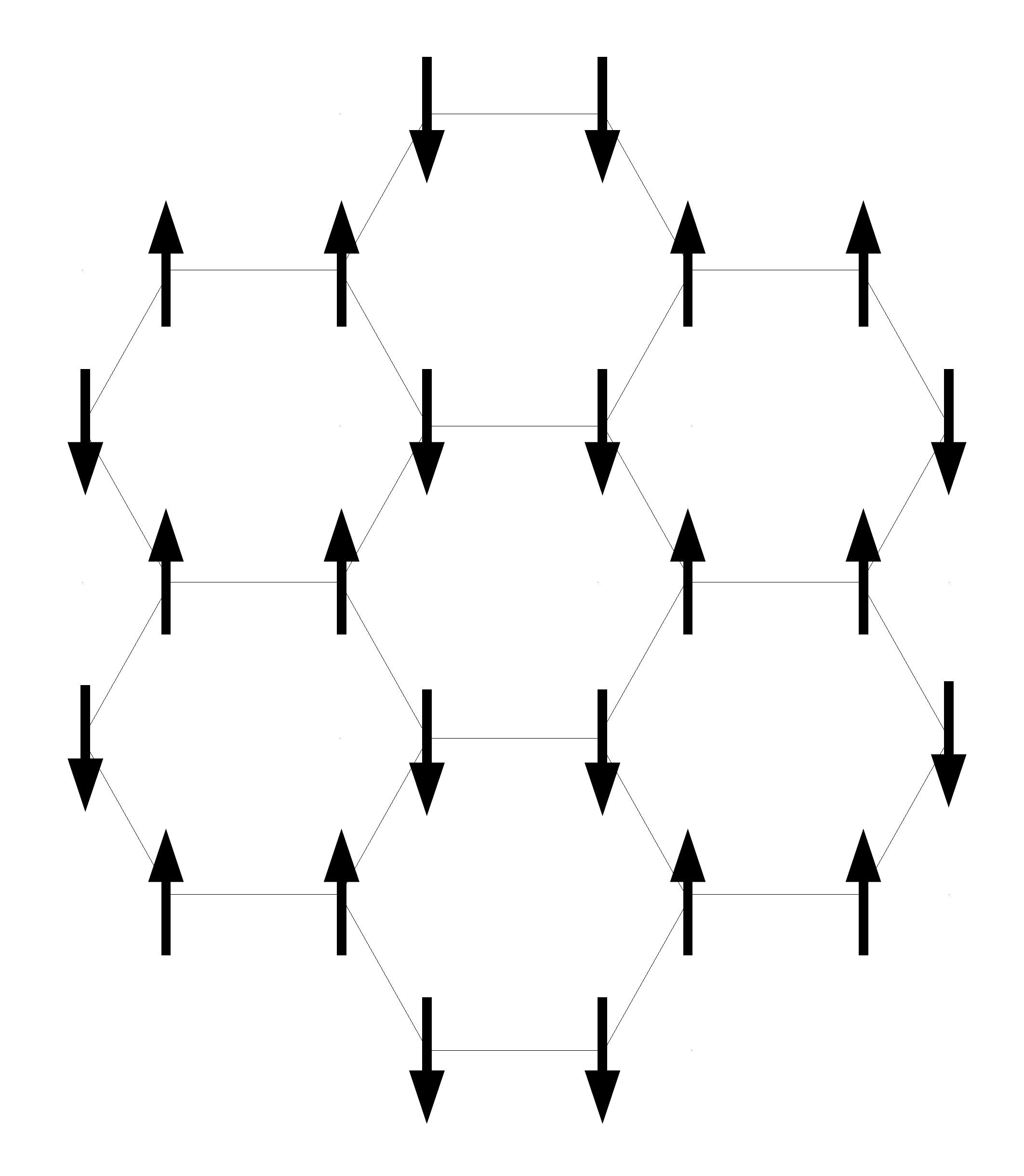}
\label{fig_stripy}
}

\caption{(a) The Phase diagram for the Heisenberg-Kitaev model and the spin pattern in the (b) Neel and (c) Stripy phases.}
\end{figure}

Subsequent to this suggestion, two honeycomb lattice compounds, Na$_2$IrO$_3$\cite{2010_singh,2011_liu,2012_singh,2012_choi,2012_feng} and Li$_2$IrO$_3$ \cite{2012_singh}, have been discovered where such a Heisenberg-Kitaev (HK) model has been suggested to capture the low energy magnetic behaviour. Meanwhile there have been intense numerical studies \cite{2012_singh,2011_jiang,2011_reuther} determining the phase diagram for the model as a function of $\alpha$, magnetic field, including further neighbour interactions\cite{2011_kimchi} and even doping\cite{2011_you,2012_hyart} (we shall not consider the effect of the further neighbour interactions or doping in this paper). These studies reveal a rich phase diagram as a function of $\alpha$ which is shown in figure \ref{fig_phasediagram}. There are three phases for $\alpha \in$ [0,1]\cite{2010_chaloupka}. (1) {\it The N\'eel Phase:} At $\alpha$ = 0 we have the antiferromagnetic Heisenberg model, which gives rise to collinear N\'eel order on the bipartite honeycomb lattice (figure \ref{fig_neel}). As $\alpha$ is increased, the N\'eel state becomes unstable at $\alpha \approx$ 0.4 to a (2) {\it Stripy} order (figure \ref{fig_stripy}). The stripy state can be seen as antiferromagnetically coupled chains which are then coupled ferromagnetically. Between $\alpha\approx 0.4-0.8$, the stripy phase is stable. Beyond $\alpha\approx 0.8$ it gives way to a (3) {\it gapless $Z_2$ spin liquid} which is continuously connected to the gapless phase of the Kitaev model (for $\alpha=1$). While the phase transition between the N\'eel and the Stripy phase appears to be discontinuous, numerical studies including density matrix renormalization group (DMRG)\cite{2011_reuther} and exact diagonalization (ED)\cite{2010_chaloupka} results suggest that the transition between the spin liquid and the stripy state is continuous or weakly first-order. DMRG also indicates that turning on a magnetic field at the critical point between the spin liquid and the stripy phase immediately opens up a polarized phase\cite{2011_reuther} and hence suggests that the phase transition between the spin liquid and the stripy phase may actually be governed by a multi-critical point.

In this paper, we explore the nature of the phase transition between the $Z_2$ spin liquid and the stripy ordered phase as a function of $\alpha$. Contrary to the original description of the spins in terms of Majorana fermions employed by Kitaev, \cite{2006_kitaev} we utilize a more conventional slave-particle approach to describe the Kitaev spin liquid which is then easily extended to include the Heisenberg term. This helps us to describe the transition between the stripy state and the spin liquid state within a slave-particle mean field theory. Our slave-particle formulation differs from that of Burnell {\it et al.}\cite{2011_burnell} and You {\it et al.}\cite{2011_you}, allowing us to extend our analysis into the magnetically ordered region by including a direct magnetic decoupling channel. Within our mean-field treatment, the transition appears to be first order with a discontinuous jump in the magnetic order parameter that is greater than that predicted by numerical calculations. \cite{2011_reuther} However, we find that this transition is brought about by subtle non-perturbative effects associated with the confinement of a gapped $U(1)$ spin liquid in two spatial dimensions.\cite{1987_polyakov} In particular, we find, within mean-field theory, that on decreasing $\alpha$ from 1 the gapless $Z_2$ spin liquid goes into a gapped $U(1)$ spin liquid phase with the simultaneous onset of magnetic order, albeit discontinuously. However, such a gapped $U(1)$ spin liquid is unstable to non-perturbative instanton effects in two spatial dimensions,\cite{1987_polyakov} which leads to immediate confinement of the spinons resulting in a conventional stripy order. We discuss the possible limitations of the present mean field theory and point out that  non-perturbative quantum fluctuations beyond mean-field may allow a more exotic continuous transition.

While, at present, it is not clear whether the HK model is relevant to the physics of the two above compounds in particular \cite{2011_bhattacharjee} (and in any case these compounds order magnetically), it presents an interesting microscopic Hamiltonian in several aspects. In addition to having a rich phase diagram of magnetically ordered and spin liquid phases, as discussed below, it offers an opportunity to study the effects of perturbation around an exactly solvable $Z_2$ spin liquid (obtained at $\alpha=1$)\cite{2011_reuther,2011_mandal,2011_tikhonov}. This latter direction allows us to study regular slave-particle mean-field theories \cite{1987_anderson,2002_wen,2011_burnell,2011_kimchi,2011_mandal,2012_hyart,2011_you} in a more controlled setting. In fact, the slave-particle mean-field theory is expected to be exact at the exactly solvable point $\alpha =1$.

The rest of the paper is organized as follows. We begin by introducing the HK model in detail in Section \ref{sec_hkmodel}, and describe a change of basis\cite{2010_chaloupka} which we will use throughout the paper. This basis change allows us to capture the transition between the stripy magnet and the spin liquid easily. We then formulate the slave particle description of the model which we use to gain insight into the Kitaev model in Section \ref{sec_slave}. We examine the exactly solvable Kitaev limit of this model within this formulation, and show that the properties of the model are reproduced within our formulation. An examination of the gauge structure of the model follows, which we use to argue that the $Z_2$ spin liquid phase is stable prior to the formation of magnetic order. In the presence of magnetic order, however, the invariant gauge group is changed into a $U(1)$ structure. In Section \ref{sec_results_mft}, we describe the mean field results in detail. We also argue that these results indicate that the transition is driven by confinement of the spinons, once we go beyond mean field theory. Finally, in Section \ref{sec_conclusion}, we conclude with a discussion of our results and indicate the possibility of continuous transition induced by quantum fluctuations. The details of mean-field self-consistent equations are given in Appendix \ref{app:MFT}.

%%%%%%%%%%%%%%%%%%%%%%%%%%%%%%%%%%%%%
\section{The Heisenberg-Kitaev Hamiltonian}
\label{sec_hkmodel}

\begin{figure}
\centering
\includegraphics[scale=0.68]{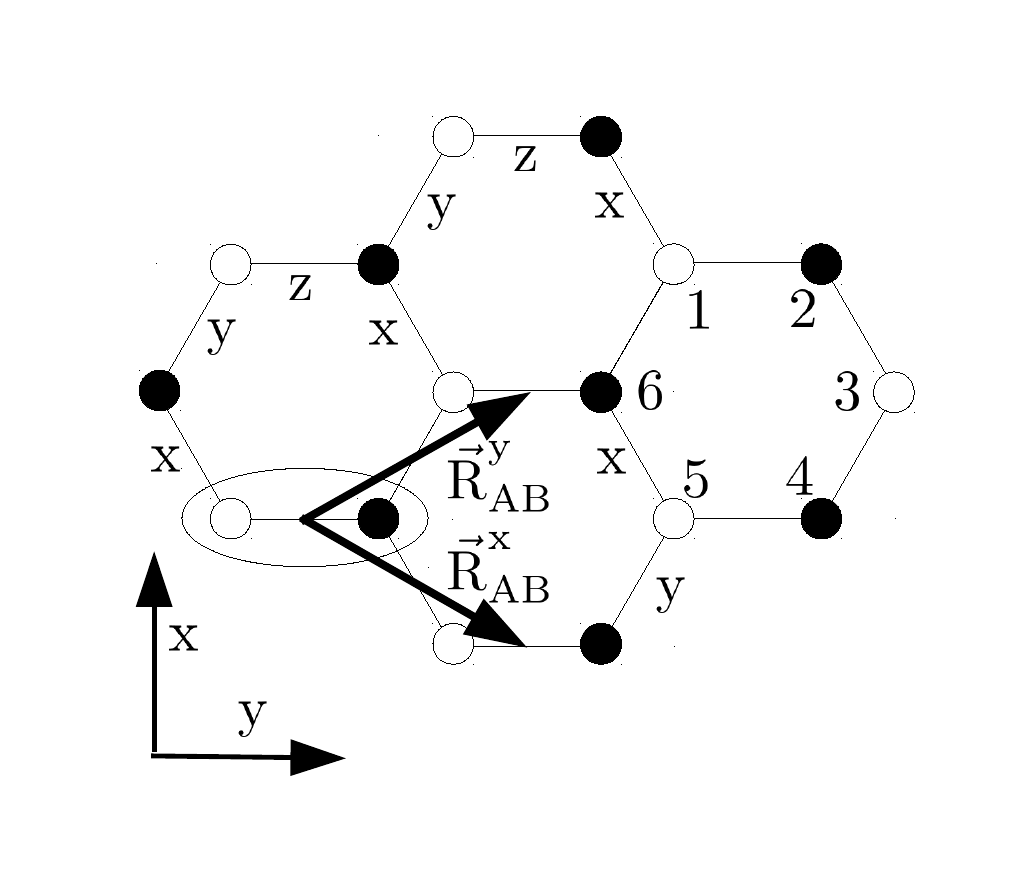}
\caption{The different interactions of the Kitaev model\cite{2006_kitaev} where the $x,y$ and the $z$ bonds are shown. The black and the white circles denote the two sublattices $A$ and $B$. ${\vec R}_{AB}^x=\frac{1}{2}\left[1,-\sqrt{3}\right]$ and ${\vec R}_{AB}^y=\frac{1}{2}\left[1,\sqrt{3}\right]$ are the two unit vectors.}
\label{fig_kitaev}
\end{figure}

We start with the discussion of the Heisenberg-Kitaev Hamiltonian. The HK Hamiltonian is given by Eq \ref{eq_hkhamiltonian}, where the Heisenberg and the Kitaev\cite{2006_kitaev} terms are given by
\begin{align}
H_{\rm H} &= \sum_{\langle ij\rangle} \vec{S}_i\cdot \vec{S}_j,\\
H_{\rm K} &= \sum_{\beta = x,y,z}~ \sum_{\langle ij\rangle,\beta -links}  S_i^\beta S_j^\beta.
\end{align}
${\vec S}_i$ denotes spin 1/2 operators defined on the sites of the Honeycomb lattice, and $\langle ij\rangle$ denotes the nearest neighbour bonds. The Heisenberg term ($H_{\rm H}$) is the usual spin rotation invariant antiferromagnetic Heisenberg Hamiltonian, coupling spins on all nearest neighbour bonds. In contrast, the Kitaev term \cite{2006_kitaev}($H_{\rm K}$) couples the $x$ components of the spins on one of the directions of bonds (referred to as $x-links$) on the honeycomb lattice, the $y$ components of spins on the $y-links$, and the $z$ components on the $z-links$, as shown in Figure \ref{fig_kitaev}. More precisely, the Kitaev model that we have written down is the isotropic Kitaev model where the couplings on the $x,y$ and $z$ links are equal.\cite{2006_kitaev}

The HK model does not have continuous spin rotation symmetry other than at the points $\alpha=0$ and $0.5$. This stems from the Kitaev part of the Hamiltonian which is devoid of continuous spin rotation symmetry. At this point it is useful to note the important symmetries of the HK model, which are
\begin{enumerate}
\item $\frac{2\pi}{3}$  spin rotation about [111] spin axis along  with $C_3$ lattice rotations about any site.

\item Inversion about any plaquette center

\item Inversion about any bond center.

\item Time reversal.

\end{enumerate}

The $C_3$ symmetry ensures that there are three different stripy phases, which we will refer to as the $x$, $y$ and $z$ stripy phases (see later). For the $\beta(=x,y,z)$ stripy phase, the spins are oriented along the $\beta$ axis, with the $\beta$ links being ordered ferromagnetically and the remaining two links ordered antiferromagnetically. Figure \ref{fig_stripy} shows one of the three possible stripy phases, namely $z$ stripy phase.

At the point $\alpha = 1$, the model can be exactly solved by transforming the spins into products of Majorana fermions, with a background of frozen $Z_2$ fluxes over plaquettes\cite{2006_kitaev}. This is a gapless $Z_2$ spin liquid, with strictly nearest neighbour spin-spin correlations.\cite{2007_baskaran} On the other hand, for $\alpha=0$ we have the pure spin rotation invariant nearest neighbour Heisenberg antiferromagnet where both numerical methods and semiclassical approaches give 2-sub-lattice Neel order.\cite{2010_mulder} In addition to these points, the model has another exactly solvable point at $\alpha=0.5$, where the stripy state is the exact ground state.\cite{2010_chaloupka,2005_khaliulin} This is easy to see by doing a selective rotation of the spins on the honeycomb lattice. It turns out that this rotated basis is useful to describe the transition between the stripy phase and the spin liquid. Hence, we shall recall the the essence of the rotation as pointed out by Khaliulin\cite{2005_khaliulin} and Chaloupka {\it et al.}\cite{2010_chaloupka}
%%%%%%%%%%%%%%
\subsection{The HK model in the rotated basis}

\begin{figure}
\centering
\includegraphics[scale=0.58]{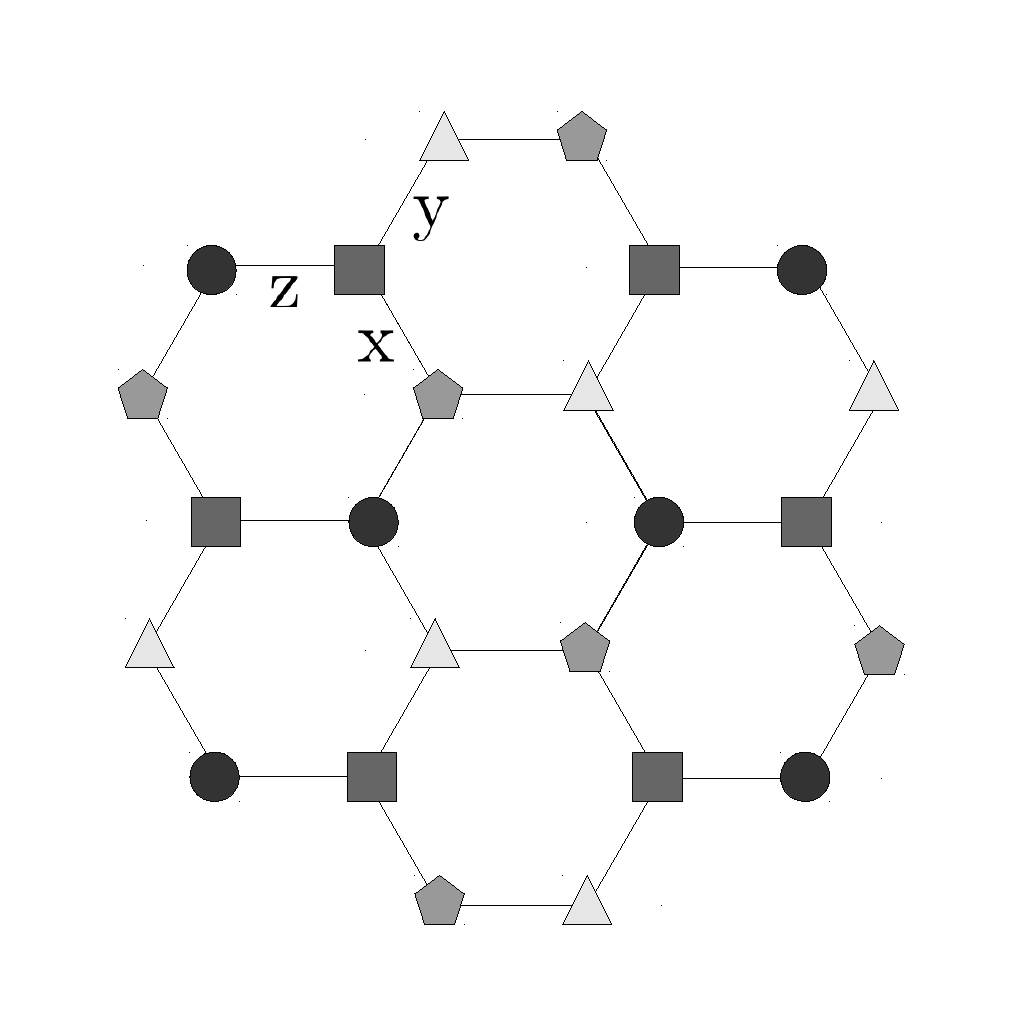}
\caption{The rotated basis: the squares are left invariant, the circles are rotated about the z bonds, the triangles about the x bonds and the pentagons about the y bonds. This rotation was first described by Khaliulin\cite{2005_khaliulin} and Chaloupka {\it et al.}\cite{2010_chaloupka}}
\label{fig_rotate}
\end{figure}

The transformation of the spin basis required to reveal the exactly solvable point at $\alpha=0.5$ is described in figure \ref{fig_rotate}.  This transformation requires different spins to be rotated about different axis, depending on their position in the lattice as described in the figure. We first choose a set of spins which are positioned on third nearest neighbour sites at opposite corners of the hexagons throughout the lattice, and hold these spins fixed. We next rotate the three spins that are adjacent to these fixed spins by $\pi$ about the spin axis corresponding to the bond which connects it to the fixed spin. This has the net effect of transforming the Heisenberg term as
\beq
H_{\rm H} \rightarrow -H_{\rm H} + 2H_{\rm K}
\eeq
and leaving the Kitaev term invariant, {\it i.e.}
\begin{align}
H_{\rm K}\rightarrow H_{\rm K}.
\end{align}
From now on, we use spins in the rotated basis. However, for the sake of brevity, we shall continue to use the same symbol for the spins and the Hamiltonians. In this basis, the Hamiltonian (given by Eq. \ref{eq_hkhamiltonian}) becomes
\beq
H\rightarrow H = -(1 - \alpha)H_{\rm H} - 4(\alpha - \frac{1}{2}) H_{\rm K}.
\eeq
In this form, the exactly solvable point at $\alpha = 0.5$ is clearly visible, as here the coefficient of the $H_K$ term is zero and this is simply the ferromagnetic Heisenberg model with a ferromagnetic ground state in terms of the rotated spins. On undoing the rotations we recover the stripy-antiferromagnetic ordering in terms of the un-rotated spins.\cite{2005_khaliulin,2010_chaloupka}

Since we wish to particularly examine the transition between the Kitaev spin liquid and the stripy anti-ferromagnetic state, we find it easier to use this rotated basis. Also, it is helpful to think about deviations from the exactly solvable point at $\alpha=0.5$ in order to simplify the couplings in the region of interest. We achieve this by introducing the parameter
\begin{align}
\delta=\alpha-\frac{1}{2}.
\end{align}
This gives
\begin{align}
H = -(\frac{1}{2} - \delta)H_{\rm H} - 4\delta H_{\rm K}.
\label{eq_finalmodel}
\end{align}
Finally, we restrict ourselves to the region $\delta \in [0,0.5]$ where  these couplings are purely ferromagnetic. We remind ourselves that $\delta=0$ now refers to the exact ferromagnetic state (or the stripy state in the original basis) while $\delta=0.5$ refers to the exactly solvable Kitaev point.

We take this rotated model as the starting point of our slave-particle analysis.

%%%%%%%%%%%%%%%%%%%%%%%%%%%%%%%%%%%%
\section{Slave Particle formulation}
\label{sec_slave}

Having written down the Hamiltonian (Eq. \ref{eq_finalmodel}) in the desired form, we now introduce the slave-particle decomposition of the (rotated) spins. We write the spin-$1/2$ operator as a bilinear of two spin-$1/2$ fermionic spinons as\cite{2002_wen,1987_anderson}

\beq
S_{j}^\mu = \frac{1}{2} f_{j\alpha}^\dag [\sigma^\mu]_{\alpha \beta}f_{j\beta},
\eeq
where $f_{j\sigma} (\sigma=\uparrow,\downarrow)$ are the fermionic spinon annihilation operators which satisfy regular fermionic anti-commutation relations. The above representation of the spin operators, along with the single fermion per site constraint
\begin{align}
f^\dagger_{i\uparrow}f_{i\uparrow}+f^\dagger_{i\downarrow}f_{i\downarrow}=1,
\label{eq_singoccupancydirac}
\end{align}
constitutes a faithful representation of the spin-$1/2$ operators.\cite{2002_wen,1987_anderson}

 The bilinear spin-spin interaction is a quartic term in the spinon operators. Within mean field theory, we now seek to decouple these quartic spinon terms into stable decoupling channels which are quadratic in terms of the spinons operators. In general, we need to keep both the particle-hole and particle-particle channels for the spinons. \cite{2002_wen}

 However, we note that both the terms in the final Hamiltonian (Eq. \ref{eq_finalmodel}) have ferromagnetic interactions. Thus the usual decoupling\cite{2002_wen} in terms of the spin-singlet particle-hole and particle-particle channels is unstable within a auxiliary field decoupling scheme. Instead, it was shown by Shindou and Momoi\cite{2009_shindou} that the correct spin liquid decoupling scheme for such interactions is into the spin triplet channels (both particle-hole and particle-particle). This is done as follows. We write the $\alpha$-th component of the spin-spin interaction as

\begin{align}
S_i^\alpha S_{i+p}^\alpha = \frac{1}{2}\sum_{\beta = x,y,z} \left(1-\delta_{\alpha,\beta}\right)\left[E^{\beta\dag}_{i,p}E^{\beta}_{i,p} + D^{\beta\dag}_{i,p}D^{\beta}_{i,p}\right] - \frac{n_i}{4},
\end{align}
where $\delta_{\alpha,\beta}=1(0)$ for $\alpha=\beta(\alpha\neq\beta)$ (not to be confused with the parameter $\delta$) is the Kronecker delta function,
\begin{align}
E^\mu_{i,p} &= \frac{1}{2} f_{i+p\alpha}^\dag [\sigma^\mu]_{\alpha \beta}f_{i\beta}, \\
D^\mu_{i,p} &= \frac{1}{2} f_{i+p\alpha}[i\sigma^y \sigma^\mu]_{\alpha \beta}f_{i\beta}
\end{align}
and $n_i=1$ is the number of spinons per site.

In addition to these hopping and pairing decouplings which capture the spin liquid, we introduce a direct channel or magnetic decoupling,
\beq
m_j = \frac{1}{2} \langle f_{j\alpha}^\dag [\sigma^z]_{\alpha \beta}f_{j\beta}\rangle,
\eeq
which, without loss of generality, we choose to be in the $S^z$ direction. We include this decoupling explicitly in order to access the ferromagnetic state and due to the fact that it is the competing order in the spin liquid phase. It is important to note that when this operator has a non-zero expectation value (in the unrotated basis) it explicitly breaks the the discrete symmetry corresponding to a lattice rotation by $\frac{2\pi}{3}$ about an individual site in conjunction with a spin rotation by $\frac{2\pi}{3}$ about the [111] spin axis (refer to our discussion of the symmetries of the HK model).

Using the above general ansatz, the mean-field spinon Hamiltonian for the rotated HK model is given by
\begin{align}
H_{HK}=-(\frac{1}{2}-\delta)H^{MF}_H-4\delta H^{MF}_K,
\label{eq_fulmfham}
\end{align}
where

\begin{widetext}
\begin{align}
H_H^{MF} = \frac{1}{4}\sum_{i}\sum_p &\Big( m(f_{i,\alpha}^\dag [\sigma^z]_{\alpha \beta} f_{i,\beta}+ f_{i+p,\alpha}^\dag [\sigma^z]_{\alpha \beta} f_{i+p,\beta}) - 2m^2 +(f_{i,\alpha}^\dag \vec{E}_{i,p}\cdot \vec{\sigma}_{\alpha \beta} f_{i+p,\beta} +h.c.)- 2|\vec{E}_{i,p}|^2 \\
&~~~~~+ (f_{i,\alpha}^\dag \vec{D}_{i,p}\cdot (-i\vec{\sigma}\sigma^y)_{\alpha \beta} f_{i+p,\beta}^\dag +h.c.) - 2|\vec{D}_{i,p}|^2\Big), \nonumber\\
H_K^{MF} = \frac{1}{4}\sum_{i}\sum_{p,r}&(1-\delta_{p,r}) \Big((f_{i,\alpha}^\dag E^r_{i,p} \sigma^r_{\alpha \beta} f_{i+p,\beta} +h.c.) - 2|E^r_{i,p}|^2 + (f_{i,\alpha}^\dag D^r_{i,p} (-i\sigma^r\sigma^y)_{\alpha \beta} f_{i+p,\beta}^\dag +h.c.) - 2|D^r_{i,p}|^2]\Big).
\label{eq_mftham}
\end{align}
\end{widetext}

 Here $i$ refers to the honeycomb lattice sites and $p,r(=x,y,z)$ correspond to the link types and spin components respectively. Eq. \ref{eq_mftham} is the most general mean field Hamiltonian with the above decoupling channels.

Taking the Fourier transform and restricting the parameters $\vec{E}$, $\vec{D}$ and $m$ for each bond type to have the symmetry of the lattice, we get
\begin{widetext}
\begin{align}
H_H^{MF} = &2\sum_k\Big((f_{k,\alpha ,A}^\dag \epsilon_{\alpha \beta}(k) f_{k,\beta ,B} +h.c.)+ (f_{k,\alpha ,A}^\dag \Delta_{\alpha \beta}(k) f_{-k,\beta ,B}^\dag +h.c.) \Big) - \frac{N_{site}}{4}\sum_p (|\vec{E}_p|^2 + |\vec{D}_p|^2)\\
&~~~~~~~~~~+2\sum_k \sum_{\eta = A,B} f_{k,\alpha ,\eta}^\dag \Omega_{\alpha \beta}f_{k,\beta ,\eta} - \frac{3N_{site}}{4}m^2, \nonumber \\
H_K^{MF} = &2\sum_k\Big((f_{k,\alpha ,A}^\dag \tilde{\epsilon}_{\alpha \beta}(k) f_{k,\beta ,B} +h.c.)+ (f_{k,\alpha ,A}^\dag \tilde{\Delta}_{\alpha \beta}(k) f_{-k,\beta ,B}^\dag +h.c.) \Big) - \frac{N_{site}}{4}\sum_{p,r}(1-\delta_{p,r}) (|E_p^r|^2 + |D_p^r|^2),
\end{align}
\end{widetext}
where $N_{site}$ is the number of lattice sites and we have defined the Fourier transform of the spinons as:
\begin{align}
f_{{\bf k},\alpha,L}=\frac{1}{\sqrt{N}}\sum_{{\bf R}_i}e^{i{\bf k}\cdot{\bf R}_i}f_{i,\alpha,L}
\end{align}
($N$ is the number of unit cells, $\alpha=\uparrow,\downarrow$ and $L=A,B$ is the sub-lattice index) and also introduced
\begin{align}
\Omega_{\alpha \beta} &= \frac{3}{8} m\sigma_{\alpha \beta}^z,\\
\epsilon_{\alpha \beta}(k) &= \frac{1}{8}\sum_p e^{i \vec{k}\cdot \vec{R}_{AB}^p}\vec{E}_p\cdot \vec{\sigma}_{\alpha \beta},\\
\tilde{\epsilon}_{\alpha \beta}(k) &= \frac{1}{8}\sum_{p,r} (1-\delta_{p,r}) e^{i \vec{k}\cdot \vec{R}_{AB}^p}E^r_p \sigma^r_{\alpha \beta},\\
\Delta_{\alpha \beta}(k) &= \frac{1}{8}\sum_p e^{i \vec{k}\cdot \vec{R}_{AB}^p}\vec{D}_p\cdot [-i\vec{\sigma}\sigma^y]_{\alpha \beta},\\
\tilde{\Delta}_{\alpha \beta}(k) &= \frac{1}{8}\sum_{p,r}(1-\delta_{p,r}) e^{i \vec{k}\cdot \vec{R}_{AB}^p}D^r_p [-i\sigma^r\sigma^y]_{\alpha \beta},
\end{align}
where we denote the unit lattice vectors (refer to figure \ref{fig_kitaev}) with
\begin{align}
\vec{R}_{AB}^x &=(\frac{1}{2},\frac{\sqrt{3}}{2}), \quad \vec{R}_{AB}^y =(\frac{-1}{2},\frac{\sqrt{3}}{2}), \quad \vec{R}_{AB}^z = 0.
\end{align}

We can write the above mean field Hamiltonian (Eq. \ref{eq_fulmfham}) in a more compact way as a Bogoliubov-de-Gennes Hamiltonian for the spinons using the 4-component Nambu spinons,
\begin{align}
\vec{f}_i^\dag &= \begin{bmatrix}
f_{i, \uparrow}^\dag & f_{i,\downarrow}^\dag & f_{i,\uparrow} & f_{i,\downarrow}
\end{bmatrix}.
\end{align}
The mean field Hamiltonian can now be written as
\beq \begin{split} \label{eq:mfhamil}
H^{MF} &= C + \sum_{i}\sum_p\big(\vec{f}_{i+p}^\dag U_{i,p} \vec{f}_{i} \\
 &~~-(\frac{1}{8} -\frac{\delta}{4})m(f_{i,\alpha}^\dag [\sigma^z]_{\alpha \beta} f_{i,\beta}+ f_{i+p,\alpha}^\dag [\sigma^z]_{\alpha \beta} f_{i+p,\beta}) \big),
\end{split}\eeq
where
\begin{align}
C = \frac{N_{site}}{4}&\big(3m^2(\frac{1}{2}-\delta ) + \sum_{p,r}[(\frac{1}{2}-\delta) +2\delta(1-\delta_{p,r})]\nonumber \\
&~~~~~~~~~~~~~~~~~~~\times(|E^r_p|^2+|D^r_p|^2)\big)
\label{eq_constant}
\end{align}
and
\begin{widetext}
\beq\begin{split}
U_{i,p} &= \frac{1}{8}\sum_r \big( -(\frac{1}{2} - \delta) - 4\delta (1-\delta_{p,r})\big) \Big(E^{r\dag}_{i,p}\sigma^r(\tau_3 + \tau_0) + E^r_{i,p}(\sigma^r)^T (\tau_3 - \tau_0)+ D^{r\dag}_{i,p} (i\sigma^y\sigma^r)\tau^- - D^r_{i,p}(i\sigma^y(\sigma^r)^T)\tau^+ \Big).
\end{split} \eeq
\end{widetext}
Here the $\sigma$ matrices are Pauli matrices operating on the spin indices and the $\tau$ matrices are Pauli matrices operating on the particle-hole indices ($\tau_0$ is the identity matrix in the particle-hole space). For the sake of clarity of notations, we have suppressed the sub-lattice index.

We now re-write the Fourier transform of the above Hamiltonian in a Nambu form to get

\begin{align}
H^{MF} = C + \sum_k \sum_{\alpha,\beta} \vec{\alpha}_{k\alpha}^\dag \tilde{H}_{k,\alpha\beta} \vec{\alpha}_{k\beta},
\label{eq_spinonhamiltonian_su2}
\end{align}
where $C$ is defined by Eq. \ref{eq_constant} and we have now used the $4$-component spinors
\begin{align}
\vec{\alpha}_{k,\beta}^\dag &= \begin{bmatrix}
f_{k,A,\beta}^\dag & f_{k,B,\beta}^\dag & f_{-k,A,\beta} & f_{-k,B,\beta}
\end{bmatrix},
\label{eq_spinonvector}
\end{align}
$A,B$ refer to the two sublattices of the honeycomb lattice (as shown in figure \ref{fig_kitaev}), $\beta=\uparrow,\downarrow$ denotes the spin and $\tilde{H}_{k,\alpha\beta}$ is given by
\begin{align}
\begin{bmatrix}
(\frac{1}{2}-\delta)\Omega_{\alpha\beta} & \xi_{\alpha\beta}(k) & 0 & \Gamma_{\alpha\beta}(k) \\
[\xi_{\beta\alpha}(k)]^\ast & (\frac{1}{2}-\delta)\Omega_{\alpha\beta} & -\Gamma_{\beta\alpha}(-k) & 0 \\
0 & -[\Gamma_{\alpha\beta}(-k)]^\ast & -(\frac{1}{2}-\delta)\Omega_{\alpha\beta} & -[\xi_{\alpha\beta}(-k)]^\ast \\
[\Gamma_{\beta\alpha}(k)]^\ast & 0 & -\xi_{\beta\alpha}(-k) & -(\frac{1}{2}-\delta)\Omega_{\alpha\beta}
\end{bmatrix}.
\label{eq_spinonmatrix}
\end{align}
We have used the following notations
\begin{align}
\xi_{\alpha \beta}(k) &= -\frac{1}{2}(1-2\delta)\epsilon_{\alpha \beta}(k) - 4\delta \tilde{\epsilon}_{\alpha \beta}(k), \\
\Gamma_{\alpha \beta}(k) &= -\frac{1}{2}(1-2\delta)\Delta_{\alpha \beta}(k) - 4\delta \tilde{\Delta}_{\alpha \beta}(k). \nonumber
\end{align}

For a given spin liquid ansatz, we diagonalize the matrix $\tilde{H}_k$ as $\rho_kD_k\rho_k^\dag$, define the vector $\vec{\gamma}_k$ = $\rho_k^\dag\vec{\alpha}_k$ and determine the values of the mean field parameters $E^\mu_p$, $D^\mu_p$ and $m_j$.

To obtain the self consistent solution, we begin with an ansatz consistent with magnetic ordering and with a combination of hopping and pairing decouplings, and allow the system to evolve to a fixed point by self-consistent iteration on the values of the mean field parameters\cite{1986_blaizot}. As all the mean field parameters are quadratic in the fermionic variables, each step in the iteration process requires an evaluation of the expectation values of quadratic fermion operators in the ground state, which are re-calculated iteratively to obtain the self-consistent solution. The details of the self-consistent equations are given in Appendix \ref{app:MFT}.

This brings us to the spin liquid ansatz which we describe next.

%%%%%%%%%%%%%%%%%%%%%%%%%%%
\subsection{The Spin Liquid Ansatz}
%%%%%%%%%
In general we have a nineteen-parameter mean field model which needs to be solved self-consistently. These fields are:
\begin{align}
\text{ On}\ p-links:\ & E^x_{i,p}, E^y_{i,p},E^{z}_{i,p}, D^x_{i,p}, D^y_{i,p},D^{z}_{i,p}
\end{align}
(where $p=x,y,z$) and the on-site magnetization $m_i$. In the spin liquid regime, the magnetization is zero and we have eighteen complex parameters and the magnetization. A self consistent mean-field analysis in terms of this eighteen(+ one) parameter model suggests that the stable mean-field states that we find involve only nine parameters or their gauge equivalent forms, in addition to magnetization in one phase. Thus we study this nine (+ magnetization) parameter model which captures both the spin liquid and the magnetically ordered ground states. The numerical calculations can further be simplified by a correct choice of gauge. To this end, we use insights from the exact solution of the Kitaev model.\cite{2006_kitaev} This, as shown below, can be obtained by choosing the following form for the nine parameters:
\begin{align}
\text{On}\ p-links:\ & D_{i,p}^x, E_{i,p}^z\in {\rm{Imaginary}},\\
& D_{i,p}^y \in {\rm{Real}},
\end{align}
($p=x,y,z$) with the remaining components set to zero. In this gauge, at the Kitaev limit, the dispersion is diagonal in terms of Majorana fermion modes\cite{2011_burnell,2011_you} as is found in the exact solution of the Kitaev model.\cite{2006_kitaev} We use the same basis as used by You {\it et al.},\cite{2011_you} in which the four Majorana fermions are defined as follows
\begin{align}
&\chi_i^0=\frac{1}{\sqrt{2}}(f_{i\uparrow} + f_{i\uparrow}^\dag );~~~~\chi_i^1=\frac{1}{i\sqrt{2}}(f_{i\downarrow} - f_{i\downarrow}^\dag )\nonumber\\
&\chi_i^2=\frac{-1}{\sqrt{2}}(f_{i\downarrow} + f_{i\downarrow}^\dag );~~~~\chi_i^3=\frac{1}{i\sqrt{2}}(f_{i\uparrow} - f_{i\uparrow}^\dag ).
\label{eq_majorana}
\end{align}

With this ansatz, we now move on to describe the two different phases and the phase transition separating them. However, before attempting to describe the general mean field results, we wish to elaborate on the Kitaev limit and the structure of the gauge theory in the next two sub-sections.
%%%%%%%%%%%%%%%%%%%%%%%%%%%%%%%%%%%%%%%%%%%%
\subsubsection{The Kitaev Limit}
In the Kitaev limit of the model, the above ansatz recovers the exact result.\cite{2006_kitaev} Most of the end results in this limit are similar to those obtained in Refs. \onlinecite{2011_burnell} and \onlinecite{2011_you}, because in this limit all these are equal to the exact solution\cite{2006_kitaev}. However, we point also point out some technical differences with our present spinon decomposition scheme. The Hamiltonian is given in terms of these Majorana fermions as:
\begin{align}
H_K &= \frac{1}{4}\sum_i\sum_{p}\Big((1-\delta_{p,z})E_{i,p}^z(\chi_i^0\chi_{i+p}^0 - \chi_i^1\chi_{i+p}^1 \nonumber\\
& ~~~~~~~~~~~~~~~~~~~~~~~~~~~~~~~~~~~~~~~~~~~~~~-\chi_i^2\chi_{i+p}^2 + \chi_i^3\chi_{i+p}^3)\nonumber\\
&-(1-\delta_{p,y})iD_{i,p}^y(\chi_i^0\chi_{i+p}^0 - \chi_i^1\chi_{i+p}^1 +\chi_i^2\chi_{i+p}^2 - \chi_i^3\chi_{i+p}^3)\nonumber\\
&+(1-\delta_{p,x})D_{i,p}^x(\chi_i^0\chi_{i+p}^0 + \chi_i^1\chi_{i+p}^1 -\chi_i^2\chi_{i+p}^2 - \chi_i^3\chi_{i+p}^3)\Big).
\end{align}

We can rewrite the single occupancy constraint for the complex fermions (eq. \ref{eq_singoccupancydirac}) in terms of the Majorana fermions\cite{2011_you} as
\begin{align}
\chi_i^0\chi_i^1\chi_i^2\chi_i^3 = \frac{1}{4}.
\label{eq_singoccupancy}
\end{align}
Using this, we can rewrite the spins in terms of the Majorana fermions as
\begin{align}
S^x_i = i\chi_i^0\chi_i^1, \qquad S^y_i = i\chi_i^0\chi_i^2, \qquad S^z_i = i\chi_i^0\chi_i^3,
\label{eq_majoranadecoupling}
\end{align}
which is the original formulation used by Kitaev in the solution of his model, with our Majorana fermions normalized such that $\{\chi_i^\alpha, \chi_j^\beta\} = \delta_{ij} \delta^{\alpha \beta}$. A set of plaquette operators,
\begin{align}
W_p = 2^6 S^x_1S^y_2S^z_3S^x_4S^y_5S^z_6,
\label{eq_fluxop}
\end{align}
are defined on the individual plaquettes of the lattice, where the sites $1-6$ traverse a honeycomb plaquette as shown in figure \ref{fig_kitaev}. (The factor of $2^6$ which is present in our definition of $W_p$ is due to the plaquette operator being written in terms of spins, rather than Pauli matrices as in the original formulation of Kitaev.\cite{2006_kitaev}) These plaquette operators commute with the original Kitaev spin Hamiltonian and with one another, which allows the Hilbert space to be split into eigen-spaces of these operators, enabling the exact solution. These operators do not commute with the mean field Hamiltonian; however,  that these operators take the same value in the mean-field solution as in the exact solution.\cite{2011_burnell}

To make a connection with Kitaev's original solution we now express our results in terms of the Majorana fermions. By construction, the Majorana fermions introduced in Eq. \ref{eq_majorana} are the modes in which the band structure is diagonal. While $\chi^1$, $\chi^2$ and $\chi^3$ form the flat bands, the single dispersing band is made up of the $\chi^0$ fermions.\cite{2011_you} In terms of the original solution of Kitaev, the dispersing fermion is the single gapless Majorana mode, while the flat band fermions describe the frozen $Z_2$ fluxes, as we now show. The flat bands arise from the fact that the mean-field Hamiltonians for $\chi^1,\chi^2$ and $\chi^3$ become disjoint, {\it i.e.}, the hopping for these fermions are non-zero only on $x,y$ or $z$ bonds respectively. For the hopping on the $z$-link, we have,
\begin{align}
\Xi (i\chi^3_i\chi^3_j-i\chi^3_j\chi^3_i)
\end{align}
where $\Xi$ is expressed in terms of the mean field parameters and $ij$ are neighbours on a z-link. The eigenvalues are given by $\pm \vert\Xi\vert$, independent of $\vec{k}$, and therefore these form the flat bands. At half filling, the lower energy state (lower flat band) is occupied. To compare with the exact solution, the Majorana bilinear $\chi^3_i\chi^3_j$ has to be identified with the $Z_2$ gauge fields defined on the $z$-links, $u^z_{ij}$.\cite{2006_kitaev} Indeed, we identify
\begin{align}
u^z_{ij}=2i\chi^3_i\chi^3_j=i(\chi^3_i\chi^3_j-\chi^3_j\chi^3_i).
\end{align}
In the ground state, clearly the eigenvalues of $u^z_{ij}$ are $\pm 1$. Similarly we can introduce $u^x_{ij}$ and $u^y_{ij}$ on $x$ and $y$ links respectively. Now we can re-write the flux operators $W_p$ in Eq. \ref{eq_fluxop} (using \ref{eq_majoranadecoupling}, \ref{eq_singoccupancy} and the fact that $\chi^\alpha_i\chi^\alpha_i=\frac{1}{2}$) as
\begin{align}
W_p &= 2^6 S^x_1S^y_2S^z_3S^x_4S^y_5S^z_6=u_{12}^zu_{23}^xu_{34}^yu_{45}^zu_{56}^xu_{61}^z
\end{align}
It is now clear that in the ground state the plaquette operators $W_p$ have an expectation value of $+1$. For a small departure from this Kitaev point, one can still use the variables $u^p_{ij}$ and $W_p$. However, these are no longer static, but acquire dynamics as the corresponding Majorana fermions starts dispersing.

The fermionic mean-field theory of this state describes a $Z_2$ spin liquid, as we will show explicitly in the next subsection. At the mean field saddle-point, the values of different parameters are given by
\begin{align}
&-iD_{i,x}^y = E_{i,x}^z = D_{i,y}^x = E_{i,y}^z = D_{i,z}^x = -iD_{i,z}^y = 0.190608i,\nonumber\\
&D_{i,x}^x = -iD_{i,y}^y = E_{i,z}^z = -0.0593918i,
\end{align}
values which have been determined by self-consistent iteration\cite{1986_blaizot}, as described above.

\begin{figure}
\centering
\includegraphics[scale=0.7]{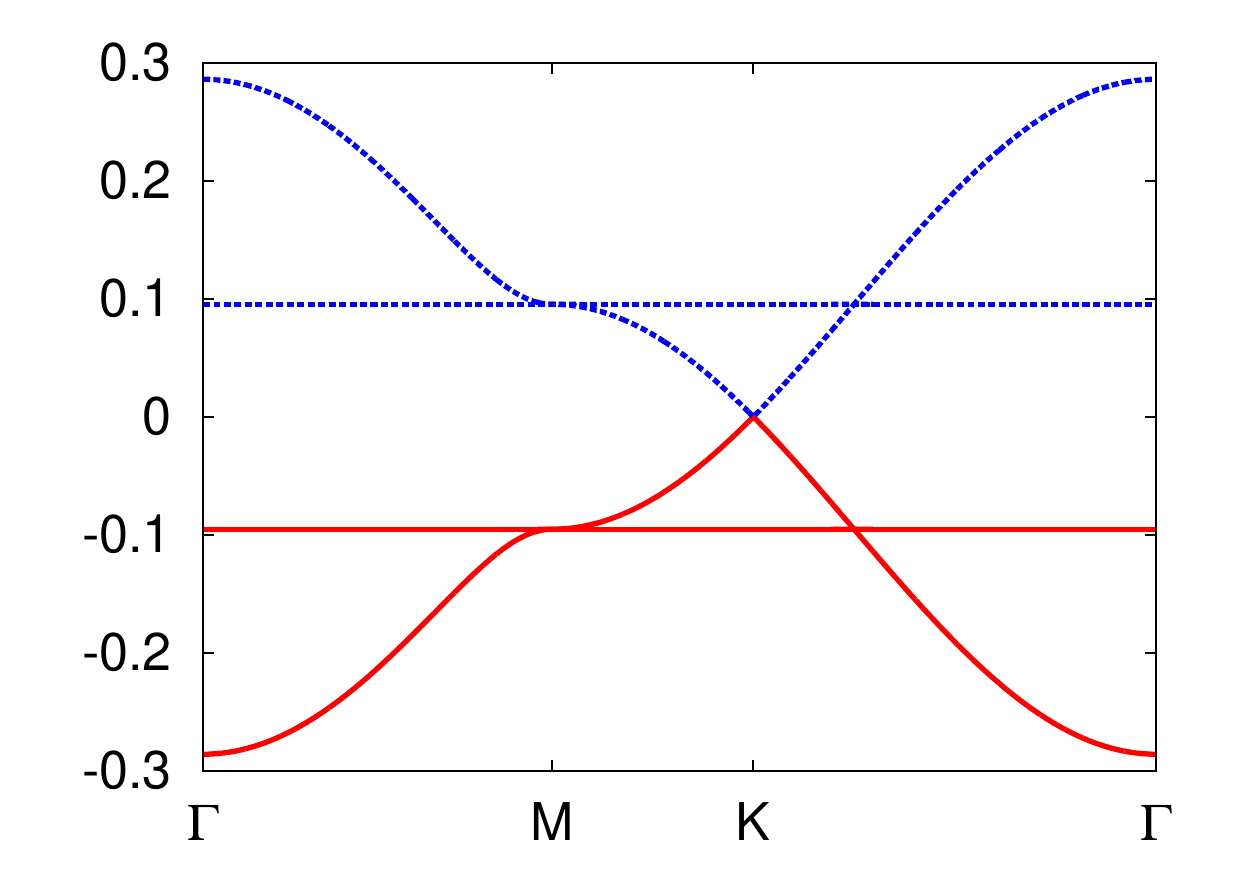}
\caption{(Color online) Spinon spectrum in the Kitaev limit. Bands shown in red are occupied, bands shown in blue are unoccupied.
\label{fig_kitaev_limit}}
\end{figure}

The resultant spinon spectrum is given in Figure \ref{fig_kitaev_limit}. There are 8 bands which, characteristic of Bogoliubov Hamiltonians, are symmetric about zero energy. The flat bands are threefold degenerate. At half filling for the spinons the lower four bands (red) are filled while the upper four bands (blue) are empty. While the flat bands are gapped, the two dispersing bands meet at the boundary of the hexagonal Brillouin zone with a characteristic Dirac spectrum. Hence the spin liquid that we are describing is indeed gapless and matches with the spinon spectrum obtained in the exact solution of the Kitaev model. This provides a useful check on the validity of our mean field solution, as well as a controlled limit from which we can perturb the model.

The presence of the pairing term indicates that, in terms of the complex fermions, the spin liquid is a ``superconductor'' for the spinons. We can analyze the symmetry of the pairing amplitude. In order to determine the properties of the pairing around the Dirac node, we isolate the dispersing band by examining the $\chi^0$ fermionic modes and returning to the original basis of Dirac fermions. For the $\chi^0$ modes, the Hamiltonian is given by
\begin{align}
H_K^0 &=\frac{M}{4} \sum_i \sum_p \chi_i^0\chi_{i+p}^0\\
&=\frac{M}{8} \sum_i \sum_p (f_{i,\uparrow} + f_{i,\uparrow}^\dag)(f_{i+p,\uparrow} + f_{i+p,\uparrow}^\dag)\nonumber\\
&=\frac{1}{8} \sum_k\sum_p( M(f_{k\uparrow A}f_{-k\uparrow B}+f_{k\uparrow A}f_{k\uparrow B}^\dag) e^{-i\vec{k}\cdot\vec{R}_{AB}^p} +h.c.)
\end{align}
where $M=0.38122i$. From here we can expand the pairing terms about the K-points in the brillouin zone. Defining $\vec{k}' = \vec{k}+(\frac{2\pi}{3},\frac{2\pi}{\sqrt{3}})$,
\begin{align}
\Delta_{dispersing}(k')&= \frac{M}{8}(1+e^{\frac{4\pi i}{3}}e^{-i\vec{k}'\cdot \vec{R}_{AB}^x} + e^{\frac{2\pi i}{3}}e^{-i\vec{k}'\cdot \vec{R}_{AB}^y})\nonumber\\
&\approx \frac{M\sqrt{3}}{16}(-k'_x+ik'_y).
\end{align}
However, we would like to emphasize that the above chiral $p$-wave pairing does not necessarily imply time-reversal symmetry breaking, which is now implemented projectively.\cite{2002_wen} The structure of the pairing terms differs from the work of Burnell and Nayak \cite{2011_burnell}, who found $p_y$ pairing about the Dirac points, by choosing a different basis for the fermions which is related to the present one by a gauge transformation.

We can further calculate the spin-spin correlation functions within mean field theory. Using the Majorana representation we find that this is given by:
\begin{align}
\langle S^\alpha_iS^\beta_j\rangle\sim\langle\chi^0_i\chi^0_j\rangle\langle\chi^\alpha_i\chi^\beta_j\rangle
\end{align}
Since the second correlation function involves absolutely flat bands, it is only non-zero when $\alpha=\beta$ and when $i=j$ or $i$ and $j$ belong to the same unit cell. Hence the spin correlation are short ranged even if the spin liquid is gapless. This is a novel feature of the Kitaev spin liquid, where exact calculations\cite{2007_baskaran} also indicate that such correlations vanish beyond nearest neighbour.

We would like to point out here that, when the model is perturbed with the Heisenberg term, the gapped flat bands acquire a weak dispersion, but still remain gapped. Within perturbation theory, this is expected to lead to exponentially decaying spin-spin correlation decaying with a length-scale characteristic of the energy-gap.\cite{2011_mandal}

%%%%%%%%%%%%%%%%%%%%%%%%%%%
\subsection{The gauge structure}

At this point, before actually discussing the results of our mean-field calculations, we wish to discuss the gauge structure of the our spin liquid ansatz.

At the outset, it should be noted that the projective symmetry group (PSG) classification including the triplet decoupling channels has not been comprehensively studied. While a comprehensive discussion of these PSGs is beyond the scope of our present work, we indicate the relevant issues in the context of the Heisenberg-Kitaev model by extending the formalism introduced by Shindou and Momoi\cite{2009_shindou}.

While the formulation outlined above is more suited to calculations of the mean field spectrum and self-consistent solutions, to decipher the nature of the spin liquid and the gauge transformations we wish to cast the above decoupling within an SU(2) formalism.

In order to examine the nature of the spin liquid state, it is worthwhile to formulate this Hamiltonian in another basis. The transformation into this basis is defined by
\begin{align}
\vec{f}_i \rightarrow \vec{f}_i'=A \vec{f}_i,\quad U_{i,p} \rightarrow A U_{i,p} A^\dag,
\end{align}
where the transformation matrix is given by
\begin{align}
A = \begin{bmatrix}
1 & 0 & 0 & 0 \\
0 & 0 & 0 & 1 \\
0 & 1 & 0 & 0 \\
0 & 0 & -1 & 0
\end{bmatrix}
\end{align}
and $\vec{f}_i$ is given by Eq. \ref{eq:mfhamil}. In the new basis, the $\vec{f}'_i$ are given by
\begin{align}
{\vec{f}'_i} {}^{\dag} = \begin{bmatrix}
f_{i, \uparrow}^\dag & f_{i,\downarrow} & f_{i,\downarrow}^\dag & -f_{i,\uparrow}
\end{bmatrix}.
\end{align}

In this basis, we can write the set of gauge transformations which leave our physical spin degrees of freedom invariant in a block diagonal form,
\begin{align}
W_i = \begin{bmatrix}
V_i & 0\\
0 & V_i
\end{bmatrix}
\end{align}
where the $V_i$ matrices form a two dimensional representation of SU(2). The spinon Hamiltonian (Eq \ref{eq_spinonhamiltonian_su2}), when written in the new basis, is  invariant under the simultaneous gauge transformation
\begin{align}
\vec{f}'_i \rightarrow W_i \vec{f}'_i,\quad U'_{i,p} \rightarrow W_{i+p} U'_{i,p} W_{i}^\dag .
\end{align}
where $U'_{i,p}=A U_{i,p}A^\dag$ gives the analog of Bogoliubov-de-Gennes Hamiltonian in the new basis.

In order to study the low energy degrees of freedom in this theory, we allow gauge fluctuations of the $U'_{i,p}$ matrices of the form
 \begin{align}
U'_{i,p} = \bar{U}'_{i,p} e^{ia_{i,p}^l \kappa^l},
\end{align}
where these $\kappa^l$ matrices are block diagonal four by four matrices
\begin{align}
\kappa^l = \begin{bmatrix}
\eta^l & 0 \\
0 & \eta^l
\end{bmatrix},
\label{eq_gaugegen}
\end{align}
where the $\eta^l$ are Pauli matrices which act on the
gauge degree of freedom, and generate our gauge transformations. We also take note of a set of matrices which generate our spin rotational symmetry, which in this basis are given by
\begin{align}
\Sigma^l = \sigma^l \otimes I,
\label{eq_spingen}
\end{align}
where the $\sigma^l$ are again Pauli matrices, acting on the spin degrees of freedom of our fermions.

To determine the gauge structure, we now consider the product of the $\bar{U}'_{i,p}$ matrices around different lattice loops based at any site $i$,\cite{2002_wen}
\begin{align}
P(C_i) = \prod_C \bar{U}'_{i,p}
\end{align}
Here the product is taken over the bonds of the loop $C_i$, beginning and ending at the site i. Using the given notations, we can write our $\bar{U}'_{i,p}$ matrices as
\begin{align}
\bar{U}'_{i,p} = \xi^{\alpha\beta}\Sigma^\alpha\kappa^\beta
\end{align}
where $\Sigma^\alpha=(\Sigma^0,\Sigma^1,\Sigma^2,\Sigma^3)$, $\kappa^\beta=(\kappa^0,\kappa^1,\kappa^2,\kappa^3)$ ($\Sigma^0$ and $\kappa^0$ are $4\times 4$ identity matrices and the other matrices are given by Eqs. \ref{eq_gaugegen} and \ref{eq_spingen}), and the $\xi^{\alpha\beta}$ are complex numbers. We note that, unlike the singlet case, in the triplet decoupling scheme both the gauge and the spin generators enter in $\bar{U}'_{i,p}$. (In other words, in the singlet decoupling\cite{2002_wen} $\Sigma^0$ is the only spin generator that enters since the decoupling channels are invariant under spin rotations).

Similarly the loop function can be written as
\begin{align}
P(C_i) =  \tilde{\xi}^{\alpha\beta}\Sigma^\alpha\kappa^\beta
\end{align}
where the $\tilde{\xi}^{\alpha\beta}$ are determined by the values of the $\xi^{\alpha\beta}$ on the links of the loop $C_i$.

If all the loop functions, based at any site, can be rotated into a form which commutes with one or more gauge generators, then the set of such generators form the invariant gauge group (IGG) and the low energy gauge fluctuations belong to the IGG.\cite{2002_wen}

Taking our ansatz in the Kitaev limit of the model, the structure of the $\bar{U}'_{i,p}$ matrices is given by

\begin{align}
\bar{U}'_{i,p} = -R\kappa^3\Sigma^3 + iQ\kappa^2\Sigma^2 + P\kappa^1\Sigma^1,
\end{align}
where
\begin{align}
R = \frac{1}{4}(1-\delta_{p,z})&E_{i,p}^{z},~~~~~~
P = \frac{1}{4}(1-\delta_{p,x})D_{i,p}^{x},\nonumber\\
Q &= \frac{1}{4}(1-\delta_{p,y})D_{i,p}^{y}.
\end{align}
The loop functions in this limit (in our choice of gauge) have a typical structure which is given by
\begin{align}
P(C_i) = T(\kappa^0\Sigma^0 - \kappa^1\Sigma^1 + \kappa^2\Sigma^2 + \kappa^3\Sigma^3),
\end{align}
where $T$ is a constant. We find that these cannot be brought into a form which commutes with any of the gauge generators, and hence the only kind of
low energy gauge fluctuation allowed has the form $W_i=e^{i\epsilon_i\kappa^l}$ where $\epsilon_i=0$ or $\pi$. This gives an IGG$\equiv Z_2$ and we have a $Z_2$ spin liquid. This spin liquid has gapless Majorana excitations (see previous sub-section) and is indeed the Kitaev spin liquid. This IGG survives throughout the entire regime of $\delta$ in which magnetic order is absent. This completes our discussion regarding the invariant gauge group of the gapless spin liquid, which we have now shown to be $Z_2$, as expected from Kitaev's exact solution.\cite{2006_kitaev}

However, in the presence of magnetic ordering the above picture is no longer true. Magnetic order drives the $E_{i,p}^z$ parameter to zero, which changes these $\bar{U}'_{i,p}$ matrices into a form given by
\begin{align}
\bar{U}'_{i,p} = i\tilde{Q}\kappa^2\Sigma^2 + \tilde{P}\kappa^1\Sigma^1,
\end{align}
where
\begin{align}
\tilde{P} &= \frac{1}{8}[(\frac{1}{2}-\delta) +4\delta(1-\delta_{p,x})]D_{i,p}^{x},\\
\tilde{Q} &= \frac{1}{8}[(\frac{1}{2}-\delta) +4\delta(1-\delta_{p,y})]D_{i,p}^{y}.
\end{align}
The loop functions are now given by
\begin{align}
P(C_i) = \tilde{T}(\kappa^0\Sigma^0 + \kappa^3\Sigma^3),
\end{align}
where $\tilde{T}$ is a different constant. It is now easy to show that the $\kappa^3$ matrix commutes with these products, and the gauge transformations of the form $W_i=e^{i\theta_i\kappa^3}~~ (\theta_i\in[0,2\pi))$ leave our ansatz invariant. Thus, in this case, the IGG$\equiv U(1)$ and hence the low energy gauge fluctuations are described by a compact $U(1)$ gauge theory which has a gapless photon and also instantons (space-time monopoles) where the gauge flux may change in integral multiples of $2\pi$. In addition, using the gauge transformation
\begin{align}
\vec{f}'_{i,B} \rightarrow \begin{bmatrix}
i\sigma^y & 0\\
0 & i\sigma^y
\end{bmatrix} \vec{f}'_{i,B},\quad \bar{U}'_{i,p} \rightarrow \begin{bmatrix}
i\sigma^y & 0\\
0 & i\sigma^y
\end{bmatrix} \bar{U}'_{i,p}
\label{eq_transu1}
\end{align}
the $D_{ij}$ can now be completely rotated into the $E_{ij}$ vectors (when $E_{i,p}$ is zero) hence explicitly showing that this is a $U(1)$ spin liquid. Later, we shall see that the spinon spectrum is gapped (in the presence of magnetic ordering), and that this state is actually a gapped $U(1)$ spin liquid which is unstable to confinement in two spatial dimensions. This significance of this instability will be discussed later.

%%%%%%%%%%%%%%%%%%%%%%%%%%%
\section{The results of the mean field theory and beyond}
\label{sec_results_mft}

\begin{figure*}
\centering
\subfigure[]{
\includegraphics[scale=.65]{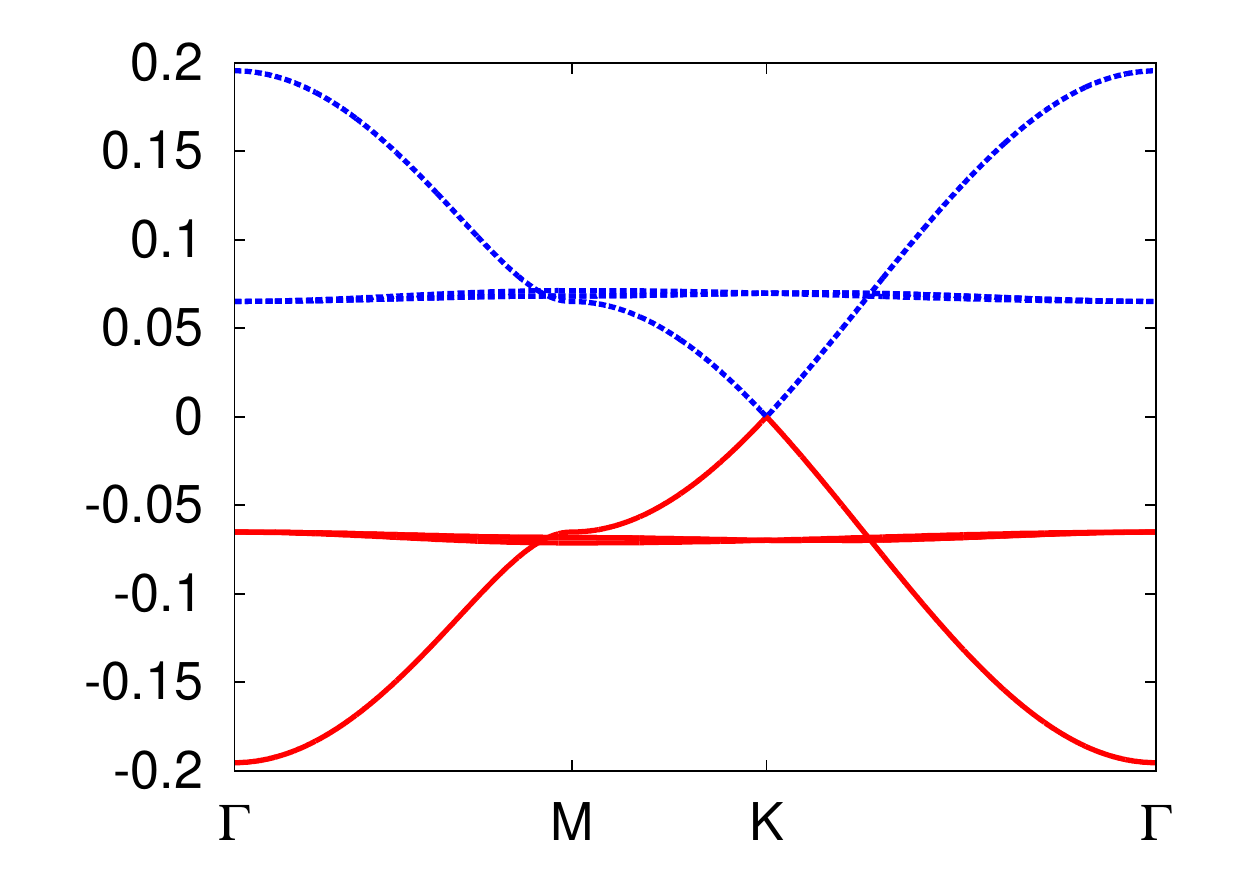}
\label{fig_kit_limit}
}
\subfigure[]{
\includegraphics[scale=.65]{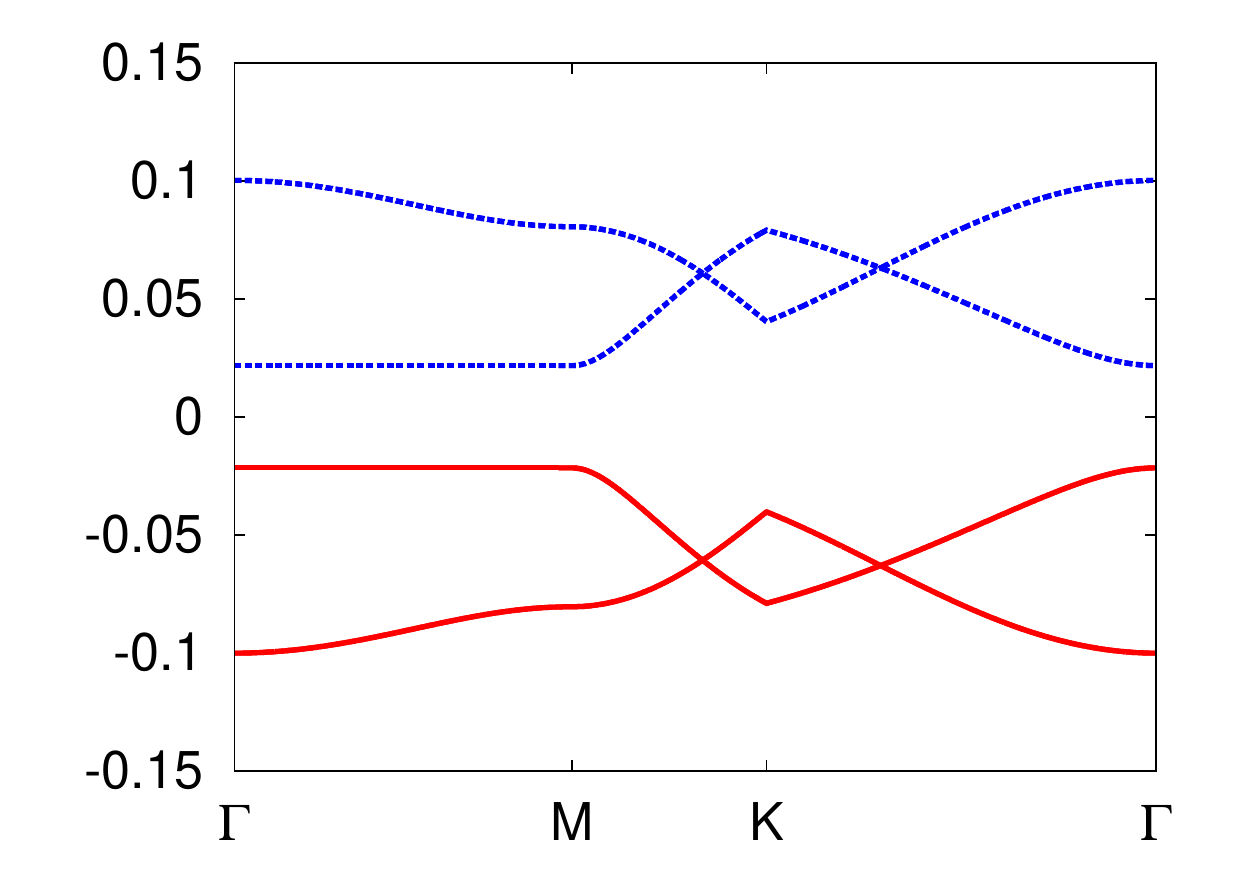}
\label{fig_mag_spinon}
}
\caption{(Color online) (a) The spinon band structure for the Heisenberg-Kitaev model with nonzero Heisenberg coupling and zero net magnetization, at the point $\delta = 0.3$. (b) The spinon band structure for the Heisenberg-Kitaev model with nonzero Heisenberg coupling and non-zero net magnetization, at the point $\delta = 0.23$.}
\end{figure*}

We now discuss the results of our mean field calculations as a function of $\delta$. These mean field results are obtained by solving the self-consistent equations (given in Appendix-\ref{app:MFT}) for the various mean field parameters.

\paragraph{Region surrounding the Kitaev limit $(\delta\approx 0.5)$: }
Near the Kitaev limit, we find that the spinon bands which were flat in the pure Kitaev limit gain a dispersion, with energy which scales with the distance from the exactly solvable point. These bands do not contribute significantly to the low energy theory due to the fact that they remain fully gapped, and the low energy spinon excitations remain consistent with those of the pure Kitaev model. Figure \ref{fig_kit_limit} shows the band structure in this region, at the value of $\delta = 0.3$. As the strength of the Heisenberg coupling is increased, the mean field parameters show only a slight change prior to the onset of magnetic ordering.

\paragraph{Region with non-zero magnetic order $(\delta\le 0.26)$: }
At $\delta \approx$ 0.26, we find that the system begins to admit a non-zero magnetic order parameter as a self-consistent solution. The magnetic order parameter jumps discontinuously to a finite value at this point, indicating a first order phase transition. The spinon band structure differs significantly in this phase, which includes the formation of a band gap. Figure \ref{fig_mag_spinon} shows a cut of the spinon bands in this region, at the value of $\delta = 0.23$. We also note that the values of all of the mean field parameters are changed by this ordering, and that the value of the hopping order parameters are driven to zero. As we continue to increase the strength of the Heisenberg coupling (decrease $\delta$) we see that the magnetic order parameter is increased, and the pairing amplitudes are driven to zero as well (below $\delta\approx 0.15$). Once the pairing amplitudes are zero, all the spinon bands become flat (not shown) and also have an energy gap.

\begin{figure}
\begin{center}
\includegraphics[scale=.7]{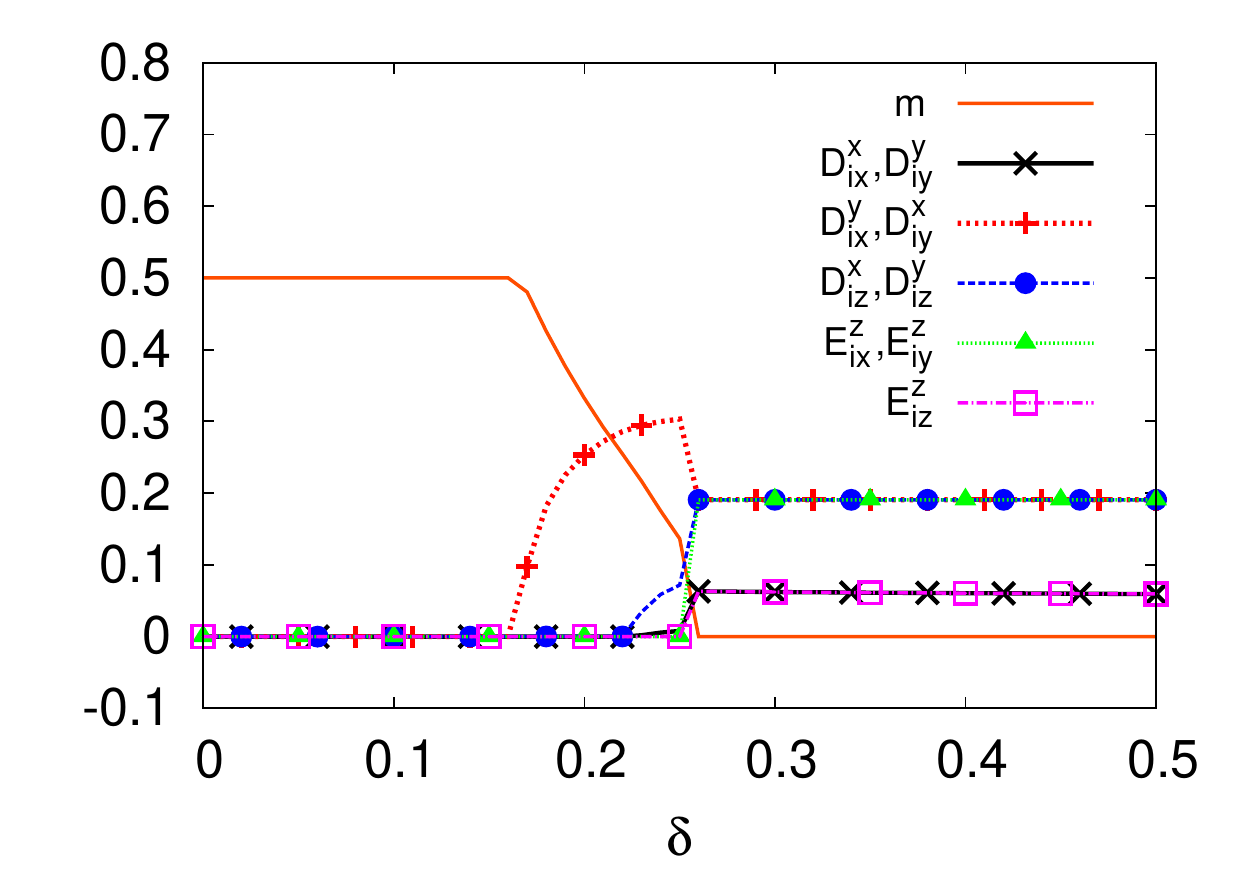}
\end{center}
\caption{(Color online) The magnitude of the mean field parameters, plotted as a function of $\delta$. The onset of magnetic order triggers a first order phase transition. The symbols are a guide for the eye.}
\label{fig_parameters}
\end{figure}

The self-consistent values of the different mean field parameters are plotted in figure \ref{fig_parameters}. This shows that the magnetic order parameter discontinuously turns on at $\delta\approx 0.26$. Below this value of $\delta$ there is finite magnetic order. Also at that value of $\delta$, $E^z_{ij}$ goes to zero discontinuously.

%%%%%%%%%%%%%%%%%%%%%%%%%%%%%
\subsection{Interpretation of mean field results}

As discussed, for $\delta\ge 0.26$ ({\it i.e.} $\alpha\ge 0.76$) there is no magnetic order and the $E_{i,p}$ and $D_{i,p}$ fields are non-zero. This, as we have already discussed already, is a $Z_2$ spin liquid which is continuously connected to the exactly solvable kitaev spin liquid (obtained for $\alpha=1$). This has gapless Majorana fermion excitations and short range spin correlations.

On the other hand, for $\delta\le 0.26$ there is magnetic order. However, at the mean field level spinon bands are well defined and there are dispersing spinons. Only the $D_{i,p}$ fields are non-zero in this phase. However, we have already shown (see Eq. \ref{eq_transu1}) that these $D_{i,p}$ fields can be gauge rotated into $E_{i,p}$ fields and hence this regime represents a $U(1)$ spin liquid. Since the spinon band structure is gapped, we essentially have a gapped $U(1)$ spin liquid with ferromagnetic order. We can call this phase a FM$^*$ in order to distinguish it with the regular ferromagnetic order. In addition to the gapped fermionic spinon excitations, there is a gapless emergent $U(1)$ gauge photon present in this phase. This arises from the underlying $U(1)$ gauge fluctuations that this phase allows. However, contrary to our regular electrodynamics, this emergent $U(1)$ gauge group is compact\cite{2002_wen}. Hence, as noted earlier, in addition to the photon, it allows instanton processes where the magnetic flux changes by integral multiple of $2\pi$. This turns out to be significant, as we discuss in the next subsection. We also point out that in general the FM$^*$ phase has a Goldstone mode (spin wave) that is in general gapped because the spin-rotation symmetry is broken explicitly (except at the points $\delta=0$ and $\delta=0.5$).

Since within mean field theory magnetic order turns on discontinuously, the transition is first order within the limits of our numerical resolution. The jump is about 20\% of the saturation value. Hence, within mean field theory, we have a first order transition between the $Z_2$ spin liquid with gapless spinons with Dirac dispersion to a FM$^*$ phase with gapped spinons.

\subsection{Beyond mean field theory: Instantons and confinement of FM$^*$ }

In this sub-section, we discuss the issue of instability of the FM$^*$ phase to a conventional ferromagnetic phase. As already pointed out, the compact $U(1)$ gauge theory that describes the low energy excitations of the FM$^*$ phase allows tunnelling processes where the magnetic flux changes (instanton events). It is known from the work of Polyakov\cite{1987_polyakov} that in $(2+1)$ dimensional compact $U(1)$ gauge theory, when the matter fields carrying electric charges(spinons) are gapped, the instanton events are always relevant. Thus, once we incorporate fluctuations to our mean field solutions, we have to take into account the effect of such tunnelling processes. Once such instanton events are taken into account, the spinons, which carry gauge charges, are confined to gauge neutral objects-- the spins. This confinement is not, however, a straightforward consequence of magnetic order, as a stable gapless U(1) phase with deconfined spinons and coexisting magnetic ordering can occur in two spatial dimensions\cite{1999_balents,2008_sung-sik,2004_hermele}. We emphasize that it is the $U(1)$ gauge structure of the magnetically ordered phase, combined with the gapped nature of the spinon excitations,\cite{1987_polyakov} which is responsible for the confinement through the proliferation of instanton events.

The above discussion indicates that once we move beyond mean field theory and take the instantons of the compact $U(1)$ gauge theory into account, the spinons in the FM$^*$ phase undergo confinement. However, the ferromagnetic order parameter would survive due to the fact that it is gauge invariant. Such a confined phase is continuously connected to the regular ferromagnetic phase for the spins and we end up with a FM phase (or the stripy phase for the unrotated spins). Thus, we indeed get a direct transition from the $Z_2$ spin liquid to a stripy phase, albeit discontinuously.
%%%%%%%%%%%%%%%%%%%%%%%%%%%%%%%%%%%
\section{Discussion}
\label{sec_conclusion}

We now summarize our results. We have obtained a slave-particle description of the HK model and used it to describe the phase transition between a spin liquid and the magnetically ordered stripy phase within slave particle mean field theory. In the Kitaev limit of the model, we have shown that this formulation reproduces the expected excitation spectrum and that the plaquette operators which enable the exact solution are in a vortex free configuration in the ground state. Upon the inclusion of a small non-zero Heisenberg term we have found a similar low energy theory, although the bands which are dispersion-free in the Kitaev limit gain a dispersion. We have analyzed the gauge structure of the model, and have seen that in the absence of magnetic order the $Z_2$ IGG which describes the Kitaev spin liquid state remains the IGG of our ansatz. The magnetically ordered phase that we get by destroying the $Z_2$ spin liquid is, within mean field theory, a gapped $U(1)$ spin liquid which has stripy magnetic order. However, existing results imply that such a spin liquid is unstable to confinement, which immediately drives a transition to the regular stripy antiferromagnetic phase. Within mean-field theory, the above transition turns out to be discontinuous. Our description allows for a coherent description of the spin liquid and the magnetically ordered phase as well as of the phase transition connecting them.

The present numerical results\cite{2010_chaloupka,2011_reuther} cannot conclusively shed light on the nature of the above transition between the spin liquid and the stripy antiferromagnet. However, these results seem to suggest that the transition is either continuous or weakly first order. While our mean field theory indicates a first order transition, we are required to incorporate quantum fluctuations beyond the mean field to address the issue of a possible continuous transition between the phases. In fact it is somewhat easy to see why the transition appears to be first order in our present calculations. Once we neglect the gauge fluctuations, we can treat the fields ${\bf E}_{ij}$ and ${\bf D}_{ij}$ as ``order parameters" along with the actual magnetization order parameter $m_i$. A Landau-Ginzburg theory in terms of these fields can then be obtained by integrating out the fermions. Such a ``multi-order parameter" Landau-Ginzburg theory with repulsive interactions between the order-parameter densities generically gives a first order transition within mean-field theory.\cite{chaikin} Hence, the results of our mean field calculations can be understood within this framework. However, a shortcoming of such a naive Landau-Ginzburg analysis is the fact that it cannot take into account the effect of gauge fluctuations. Once such fluctuations are accounted for, the ${\bf E}_{ij}$ and ${\bf D}_{ij}$ fields can no longer be treated as order parameters, since they are not gauge invariant, and so the above naive Landau-Ginzburg theory breaks down. This opens up the possibility of subtle gauge fluctuation effects driving this transition to second order. It is known from earlier studies in frustrated magnets that such ``Landau forbidden" generic continuum quantum phase transitions may occur ({\it e.g.} deconfined quantum critical points\cite{2004_senthil}) where naive mean field considerations break down or do not apply, since there are no local order parameters ({\it e.g.} Topological phase transitions\cite{2009_kou}). Hence such a second order transition would be unconventional and potentially interesting, particularly in context of the the possibility of realizing the HK model in material systems.\cite{2010_chaloupka,2010_singh} Hence, this would be an ideal subject of a future study. Similar transitions in spin rotation invariant systems have been recently studied both numerically\cite{2010_meng,2011_jiang2} and from the field theoretical perspective.\cite{2012_moon}

\begin{acknowledgements}
We acknowledge T. Dodds, F. Burnell and S. Trebst for useful discussion. This research was supported by the NSERC, CIFAR, and Centre for Quantum Materials at the University of Toronto.
\end{acknowledgements}
\appendix

\section{Self-consistent Mean Field Theory}
\label{app:MFT}

In order to determine the values of our mean field parameters, we consider the matrix $\tilde{H}_{k}$, defined as
\begin{align}
\tilde{H}_k = \begin{bmatrix}
H_{k, \uparrow \uparrow} & H_{k, \uparrow \downarrow} \\
H_{k, \downarrow \uparrow} & H_{k, \downarrow \downarrow}
\end{bmatrix},
\end{align}
where $H_{k,\alpha \beta}$ is given by \ref{eq_spinonmatrix}. We diagonalize $\tilde{H_k}$ as U$_k$D$_k$U$_k^\dag$ and define the vector $\vec{\gamma}_k$ = U$_k^\dag\vec{\alpha}_k$, where $\vec{\alpha}_k$ is given by
\begin{align}
\vec{\alpha}^\dag_k = \begin{bmatrix}
\vec{\alpha}^\dag_{k, \uparrow} & \vec{\alpha}^\dag_{k, \downarrow}
\end{bmatrix},
\end{align}
where $\vec{\alpha}^\dag_{k,\alpha}$ is given by \ref{eq_spinonvector}. For a given value of the mean field parameters we evaluate these U$_k$ for each wavevector $k$, and determine the values of the mean field parameters $E_\mu(p)$ and $D_\mu(p)$ in this state. To do this, we must look at expectation values of the form $\langle f^\dag_{j\alpha}f_{i\beta}\rangle$ and $\langle f_{j\alpha}f_{i\beta}\rangle$. We have assumed previously that the parameters $E$ and $D$ are uniform throughout the lattice, and we use this fact to rewrite these expectation values as
\beq \begin{split}
-\langle f_{i,\beta} & f^\dag_{i+\gamma,\alpha}\rangle = -\frac{2}{N_{site}} \sum_{\gamma -links} \langle f_{i,\beta}f^\dag_{i+\gamma,\alpha}\rangle \\
&= -\frac{2}{N_{site}} \sum_{k} \langle f_{k,\beta ,A}f^\dag_{k,\alpha ,B} \rangle e^{-i\vec{k} \cdot \vec{R}^p_{AB}} \\
&= -\frac{2}{N_{site}} \sum_k \sum_{l'} \sum_{m'} \langle U_{k,l,l'}\gamma_{k,l'} U^\ast_{k,m,m'}\gamma^\dag_{k,m'} \rangle e^{-i\vec{k} \cdot \vec{R}^p_{AB}} \\
&= -\frac{2}{N_{site}} \sum_k \sum_{l' (unoccupied)} U_{k,l,l'} U^\ast_{k,m,l'} e^{-i\vec{k} \cdot \vec{R}^p_{AB}}
\end{split} \eeq
for l corresponding to $\beta ,A$, m corresponding to $\alpha ,B$, and the sum over unoccupied $l'$ referring to the eigenvectors of $\tilde{H}_k$ corresponding to positive eigenvalues. Similarly, \hspace{100 mm}
\beq \begin{split}
-\langle f_{i,\beta} & f_{i+\gamma,\alpha}\rangle = -\frac{2}{N_{site}} \sum_{\gamma -links} \langle f_{i,\beta}f_{i+\gamma,\alpha}\rangle \\
&= -\frac{2}{N_{site}} \sum_{k} \langle f_{k,\beta ,A}f_{-k,\alpha ,B} \rangle e^{-i\vec{k} \cdot \vec{R}^p_{AB}} \\
&= -\frac{2}{N_{site}} \sum_k \sum_{l'} \sum_{m'} \langle U_{k,l,l'}\gamma_{k,l'} U^\ast_{k,m,m'}\gamma^\dag_{k,m'} \rangle e^{-i\vec{k} \cdot \vec{R}^p_{AB}} \\
&= -\frac{2}{N_{site}} \sum_k \sum_{l' (unoccupied)} U_{k,l,l'} U^\ast_{k,m,l'} e^{-i\vec{k} \cdot \vec{R}^p_{AB}}
\end{split} \eeq
for l corresponding to $\beta ,A$ and m corresponding to $\alpha ,B$ on the -k portion of the vector. We also need to evaluate terms of the form $\langle f^\dag_{i\alpha}f_{i\alpha}\rangle$ to determine our magnetization m. As above, we determine these by evaluating
\beq \begin{split}
\langle f_{i,\alpha}^\dag f_{i,\alpha} \rangle &= \frac{2}{N_{site}} \sum_i \langle f_{i,\alpha}^\dag f_{i,\alpha} \rangle \\
&= \frac{2}{N_{site}} \sum_k \langle f_{k\alpha ,A}^\dag f_{k\alpha ,A} \rangle \\
&= \frac{2}{N_{site}} \sum_k \sum_{l'} \sum_{m'} \langle U^\ast_{k,l,l'} \gamma^\dag_{k,l'} U_{k,l,m'}\gamma_{k,m'} \rangle \\
&= \frac{2}{N_{site}}\sum_k \sum_{l' (occupied)} |U_{k,l,l'}|^2
\end{split} \eeq
where l corresponds to $\alpha ,A$. We can determine the occupancy of the sites on sublattice B similarily. Using these expressions, we can determine the expectation value of our mean field parameters beginning with any ansatz. Doing this iteratively (with the previous solutions as our new ansatz at each step) allows us to determine an approximate self-consistent solution to our mean-field equations.
%%%%%%%%%%%%%%%%%%%%%%%%%%%%%%%%%%%

%%%%%%%%%%%%%%%%%%%%%%%%%%%%%
\end{document}